\documentclass[sigconf,natbib=true]{acmart}

\usepackage{amsmath}

\usepackage{amssymb}
\usepackage{array}
\usepackage{booktabs}
\usepackage{tabularx}
\usepackage[inline]{enumitem}
\usepackage{graphicx}
\usepackage{subcaption}
\usepackage{hyperref}
\usepackage{xcolor}
\usepackage{tikz}
\usetikzlibrary{arrows.meta,positioning,fit,calc}
\usepackage{listings}

\usepackage{xspace}
\usepackage{etoolbox}

\newif\ifarxiv \arxivfalse
\newcommand{\fullref}[1]{\ifarxiv the full version\else Section~\ref{#1}\fi}
\newcommand{\fullrefs}[2]{\ifarxiv the full version\else Sections~\ref{#1}--\ref{#2}\fi}
\newcommand{\psecref}[1]{\ifarxiv\else\ (Section~\ref{#1})\fi}

\newcommand{\ty}[1]{\textsf{\small #1}}     %
\newcommand{\fn}[1]{\texttt{\small #1}}     %
\newcommand{\law}[1]{\textbf{#1}}           %

\newcommand{\rtag}[1]{\textbf{R#1}}   %
\newcommand{\ntag}[1]{\textbf{N#1}}   %
\newcommand{\ttag}[1]{\textbf{T#1}}
\newcommand{\ctag}[1]{\textbf{C#1}}

\DeclareRobustCommand{\revised}[2]{%
  \ifstrempty{#1}{}{{\color{orange!85!black}\textsf{\small[\,$\triangleright$~\textit{#1}\,]}}\,}%
  {\color{blue}#2}%
}

\arxivtrue
\renewcommand{\revised}[2]{#2}

\graphicspath{{figures/}}

\renewcommand\footnotetextcopyrightpermission[1]{}
\newcommand{\yes}{\textbf{\checkmark}}
\newcommand{\pt}{\ensuremath{\circ}}      %
\newcommand{\no}{\ensuremath{\times}}
\newcommand{\na}{\,--\,}
\newcommand{\rh}[1]{\rotatebox{90}{#1}}

\title{Confining Nondeterminism: AI-Driven Research Systems as DBMSs for Reliable, Non-Wasteful, Transparent, \\ and Collaborative Research [Vision]}

\author{Kyoungmin Kim}
\affiliation{\institution{EPFL}\city{Lausanne}\country{Switzerland}}
\email{kyoung-min.kim@epfl.ch}

\author{Anastasia Ailamaki}
\affiliation{\institution{EPFL}\city{Lausanne}\country{Switzerland}}
\email{anastasia.ailamaki@epfl.ch}

\begin{abstract}
{%
LLM agents that conduct research (proposing ideas, writing and
running code, analyzing results)
{can already carry a study
from research question to figures, yet cannot be fully trusted.} The
same question asked twice in a row returns different answers;
{the agent announces a number that no execution
produced, and having tools does not prevent this, because nothing
binds what the agent reports to what its tools returned}; a
{small upstream change
leaves downstream results silently stale, with no way to list which
ones}; and {the agent re-runs preprocessing
and re-writes code it has already produced, because nothing tells
it that an identical step has already run}, at the researcher's
expense. We argue these failures share one root: every step of
today's agent loop is a stochastic LLM call whose internal state
nobody, including the agent, can check. Rather
than trying to see inside the LLM, we take a lesson from
databases, which earn trust without being watched, because
deterministic operators over well-defined state make their
guarantees hold by construction. We propose organizing a research
project the same way. The project itself is a deterministic,
versioned dataflow engine (in effect, a query plan over materialized
views), and the LLM, together with the user, is a stochastic
compiler that may only \emph{edit} that plan. The executor never
calls the {LLM; LLM} output enters only as versioned code and
data that the executor then runs; and {any result the agent asserts} enters the {project's stored record}
only with an execution behind it. Five {design rules} at this boundary turn
familiar database machinery into guarantees for research.
Versioning, idempotence, rollback, provenance, evidence-backed
results, and isolation make the system \emph{reliable};
incremental maintenance, memoization, cost-based scheduling, and
approximation make it \emph{non-wasteful}; an explicit, queryable plan makes it
\emph{transparent}; and because the whole project is now a small
database instance, sharing it makes research \emph{collaborative},
down to peer review conducted as read-only queries over a submitted
instance. We map each recurring pain of research agents to the
database technique that removes it, show that prior
agent--data-systems work covers only the performance side of this
map{\ifarxiv.
This report presents the diagnosis, the requirements, and the
design; the guarantee walkthrough, a prototype with early results,
and the research agenda appear in the full version, in
preparation\else, and report early results from a prototype\fi}. The LLM, we argue,
should be the query compiler, never the executor.}
\end{abstract}


\begin{document}
\maketitle

\section{Introduction}
\label{sec:intro}

{%
A new kind of system has begun to do research automatically. Given a goal, say
``find out whether data augmentation improves this classifier,'' an
LLM-based agent proposes an approach, writes the code, runs the
experiments, reads the results, and
iterates~\cite{luAIScientist,gottweisCoScientist,russoDeepResearch}.
We call such a system an \emph{AI-Driven Research System
(ADRS)}.\footnote{Not to be confused with Berkeley's ``AI-Driven
Research \emph{for} Systems''~\cite{berkeleyADRS}, which applies AI
to systems-optimization problems. Our subject is the research system
itself.} The database community has started to take these agents
seriously as a workload; recent work redesigns data systems so that
agents can probe them efficiently~\cite{liuAgentFirst,eckmannQCP}.
{This paper asks the opposite question: if we want to trust what the agent produces, and afford to run it, what should the \emph{agent}
look like? Our answer is that it should look like a database
system.}}

{%
\textbf{Two ways they disappoint.} The first is that you cannot
trust them. Ask the same question twice
{in a row, with nothing
changed in between, and the answer changes: today's agents have no
notion of ``the same request,'' so each ask is compiled and answered
afresh.} %
{A long run ends with the agent
announcing that augmentation helps, at 94\% accuracy, when no
evaluation ever ran. Tool use does not close this gap: the agent
may execute an evaluation and still report a different figure,
or skip the run and answer from expectation, because nothing binds
a reported number to an execution.} A step that went wrong
cannot be undone, because nothing records what depended on it.
{Rolling back by hand to last week's better-performing setup rarely recovers it exactly, and nothing authoritative records which version actually scored better.}
None
of this is anecdotal: analytics agents ``take shortcuts'' and
``terminate prematurely''~\cite{russoDeepResearch}, and a recent
study catalogs dozens of such failure modes~\cite{cemriFail}.}

{%
The second is quieter: they burn money and attention. The
conversation context only grows, and the agent
{re-runs preprocessing it has already run, rewrites code
it has already written, and re-sends the same file contents as
fresh tokens on every turn.} One measurement of agents solving data
tasks found fewer than 10--20\% of agent-generated sub-plans to be
{unique; the rest were near-duplicates, issued
because the agent explores by trial and error and cannot tell what
it has already tried}~\cite{liuAgentFirst}. Meanwhile the
researcher babysits: idle while the loop grinds, then interrupted
at dawn for a decision that could have waited until office hours, %
{and experiment servers sit idle when she does not. And when results later prove unreliable, correcting them by hand can take longer than doing the work herself would have.}}

{%
\textbf{One root cause.} Both families share a root. Every step of
today's agent loop is a call to an LLM whose internal state is
hidden. What we observe is the text it emits, and the text is an
unreliable \emph{report} of what the LLM actually did or knows.
This opacity is a property of the medium, not a bug anyone can
patch. Its consequence is that
{verification never happens by default. One can
hand-write or {LLM-generate} a checker for any single step, but the architecture has
no place where checked facts live {(agent context is volatile)}, no way to tell them apart from
unchecked prose, and nothing that forces the next step to consult
them. Checking is doable, but nothing does it systematically.} So the
agent acts on claims that were never verified (the unreliability)
and re-derives facts it has already established (the waste).}

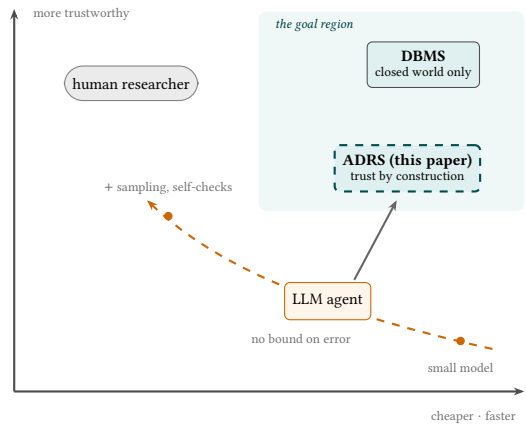
\begin{figure}[t]
\centering
\begin{tikzpicture}[x=1pt,y=1pt,
  every node/.style={font=\scriptsize, align=center},
  ctl/.style={draw=black!70, fill=teal!12, rounded corners=2pt,
              minimum height=14pt, inner sep=3pt},
  sto/.style={draw=orange!80!black, fill=orange!8, rounded corners=2pt,
              minimum height=14pt, inner sep=3pt},
  dat/.style={draw=black!50, fill=black!8, rounded corners=7pt,
              minimum height=13pt, inner sep=3pt},
  ann/.style={font=\tiny, text=black!55},
  arr/.style={-{Stealth[length=4pt]}, black!60, thick}]

\fill[teal!6, rounded corners=3pt] (134,92) rectangle (234,167);
\node[font=\tiny\itshape, text=teal!50!black, anchor=west] at (137,162) {the goal region};

\draw[arr, black!70] (42,24) -- (234,24);
\draw[arr, black!70] (42,24) -- (42,168);
\node[ann, anchor=north east] at (234,20) {cheaper $\cdot$ faster};
\node[ann, anchor=west] at (46,166) {more trustworthy};

\draw[orange!80!black, thick, dashed, -{Stealth[length=4pt]}]
  (222,40) .. controls (150,56) and (115,74) .. (92,96);
\fill[orange!80!black] (210,43) circle (1.6pt);
\node[ann, anchor=north] at (210,38) {small model};
\fill[orange!80!black] (100,90) circle (1.6pt);
\node[ann, anchor=south] at (100,95) {$+$ sampling, self-checks};
\node[ann] at (150,44) {no bound on error};

\node[dat] (hum) at (86,140) {human researcher};
\node[ctl] (db) at (196,147) {\textbf{DBMS}\\[-1pt]{\tiny closed world only}};
\node[sto] (llm) at (160,58) {LLM agent};

\node[draw=teal!60!black, thick, dashed, rounded corners=2pt, fill=teal!12,
      inner sep=3pt] (tgt) at (190,108)
  {\textbf{ADRS (this paper)}\\[-1pt]{\tiny trust by construction}};
\draw[arr] (170,66) -- (186,96);

\end{tikzpicture}
\caption{{The corner an ADRS should occupy. A DBMS is trusted at
machine cost, but only over the closed world its schema captures; a
human researcher is trusted over open-ended work, at human cost; an
LLM agent brings machine cost to open-ended work, but more compute
only raises the odds of being right, bounding no error. We aim at
the empty corner and argue it is reached the way databases reached
theirs, by construction
(Sections~\ref{sec:desiderata}--\ref{sec:model}).}}
\label{fig:quadrant}
\end{figure}

{%
\textbf{Why databases?} One response to opacity is to try to see
inside the LLM. We take a different lesson, from a system people
trust \emph{without} watching it. Nobody inspects a buffer pool to
convince themselves a transaction committed; we trust the DBMS
because its guarantees hold by construction. Its operators are
deterministic functions over well-defined state, and given that,
reproducibility, recovery, and safe concurrency are properties of
the design, not promises of good behavior.
{Figure~\ref{fig:quadrant} puts the goal on a map. The DBMS earns
this trust at machine cost, but only over the closed world its
schema captures. A human researcher is trusted over open-ended
work, but is slow, expensive, and does not scale. Today's agents
bring machine cost to open-ended work, but no amount of extra
sampling or self-checking bounds their error: spending more moves
them along a curve that never reaches the trusted region. What
research automation needs is the empty corner, the trust of the
DBMS at the cost of the agent, over the open-ended work of
research; the question this paper pursues is whether that corner
can be reached the way databases reached theirs, by construction,
even though one component remains an opaque, stochastic model.}}

{%
\textbf{Thesis: confine the nondeterminism.} It can, if the LLM is
put in the right place. A DBMS already contains an error-prone,
heuristic component, the query optimizer, and survives it by
confinement:
{every plan the optimizer may pick is an equivalent rewriting of the same query, so a bad pick returns the same answer more slowly. The optimizer decides how to compute, never what is true.} 
We propose the same
structure for research (Figure~\ref{fig:arch}). The project lives
in a \emph{deterministic, versioned dataflow engine} (\textbf{L1}):
a DAG of operators over immutable, content-addressed artifacts; in
database terms, a query plan whose intermediate results are
materialized views. The LLM, together with the user, forms a
\emph{stochastic compiler} (\textbf{L2})
{that compiles natural-language requests the
way a text-to-SQL interface compiles a question into a query,
except that the output is an \emph{edit} of the standing plan
rather than a one-off query.} One invariant separates the strata:
\textbf{L1 never calls an LLM}.
{LLM} output enters L1 only as a
concrete, versioned artifact; the LLM writes \fn{aug.py}, and a
deterministic node \emph{runs} it. Under this discipline,
{five design rules (Section~\ref{sec:model}) fix how the
two strata may interact, and every guarantee in this paper follows
from those rules rather than from the LLM behaving well.}
{This
asymmetry is exactly what confinement buys, and it is the claim
the rest of the paper cashes out: the LLM may still be wrong, but
only about what to try next, never about what was computed.} The
LLM is the query compiler, never the executor.}

\begin{figure*}[t]
\centering
\begin{tikzpicture}[x=1pt,y=1pt,
  every node/.style={font=\scriptsize, align=center},
  ctl/.style={draw=black!70, fill=teal!12, rounded corners=2pt,
              minimum height=14pt, inner sep=3pt},
  sto/.style={draw=orange!80!black, fill=orange!8, rounded corners=2pt,
              minimum height=14pt, inner sep=3pt},
  dat/.style={draw=black!50, fill=black!8, rounded corners=7pt,
              minimum height=13pt, inner sep=3pt},
  scp/.style={font=\scriptsize\itshape, text=black!55},
  arr/.style={-{Stealth[length=4pt]}, black!60, thick},
  darr/.style={-{Stealth[length=4pt]}, black!60, thick, dashed}]

\foreach \x in {0,91,182,273} {
  \draw[black!20, rounded corners=3pt] (\x,0) rectangle (\x+86,112); }
\draw[black!20, rounded corners=3pt] (364,0) rectangle (472,112);

\node[font=\scriptsize\bfseries] at (43,120) {(a) Text-to-SQL};
\node[ctl] (a1) at (43,96) {script (fixed)};
\node[sto] (a2) at (43,60) {LLM};
\node[ctl] (a3) at (43,26) {DBMS};
\draw[darr] (a1) -- (a2);
\draw[arr]  (a2) -- node[right, font=\tiny]{SQL} (a3);
\node[scp] at (43,8) {scope: a query};

\node[font=\scriptsize\bfseries] at (134,120) {(b) RAG};
\node[ctl] (b1) at (134,96) {script (fixed)};
\node[dat] (b2) at (134,60) {corpus / DB};
\node[sto] (b3) at (134,26) {LLM answers};
\draw[darr] (b1) -- (b2);
\draw[arr]  (b2) -- (b3);
\node[scp] at (134,8) {scope: a question};

\node[font=\scriptsize\bfseries] at (225,120) {(c) Semantic operators};
\node[ctl] (c1) at (225,96) {optimizer};
\node[ctl] (c2) at (203,52) {rel op};
\node[sto] (c3) at (249,52) {sem op};
\draw[darr] (c1) -- (c2);
\draw[darr] (c1) -- (c3);
\draw[arr]  (c2) -- (c3);
\node[scp] at (225,8) {scope: a query};

\node[font=\scriptsize\bfseries] at (316,120) {(d) Agent frameworks};
\node[sto] (d1) at (316,92) {LLM agent\\[-1pt]{\tiny control plane}};
\node[ctl] (d2) at (316,42) {DB $\cdot$ tools};
\draw[darr] ([xshift=-10pt]d1.south) -- ([xshift=-10pt]d2.north);
\draw[arr]  ([xshift=10pt]d2.north)  -- ([xshift=10pt]d1.south);
\node[scp] at (316,8) {scope: a task};

\node[font=\scriptsize\bfseries] at (418,120) {(e) Ours: plan compiler};
\node[sto] (e1) at (418,96) {compiler (LLM + user)};
\draw[draw=black!60, thick, rounded corners=3pt, fill=teal!6]
  (374,16) rectangle (462,64);
\node at (418,52) {\textbf{L1}: plan + executor};
\node[font=\tiny, text=black!55] at (418,40) {deterministic, versioned};
\node[ctl, inner sep=2pt, font=\tiny] (e2) at (398,26) {op};
\node[ctl, inner sep=2pt, font=\tiny] (e3) at (438,26) {op};
\draw[arr] (e2) -- (e3);
\draw[arr, very thick] (406,88) -- (406,66);
\draw[black!70, very thick] (398,78) -- (414,78);
\node[font=\tiny, anchor=east] at (403,77) {edits only};
\draw[arr] (430,66) -- (430,88);
\node[font=\tiny, anchor=west] at (432,77) {results};
\node[scp] at (418,8) {scope: the project};

\end{tikzpicture}
\caption{{Where the stochastic call sits in five
LLM--database couplings. Orange = stochastic (the LLM), teal =
deterministic, gray pills = data; solid arrows carry data, dashed
arrows are run-time control. In (a) and (b) control is
deterministic but \emph{fixed}: reliable because it decides
nothing. In (c) a deterministic optimizer adapts, but the
stochastic call sits on the data path, inside the executor, and
nothing outlives the query. In (d) the stochastic component owns
control and grades itself. In (e) the LLM authors the plan through
a gate that admits edits only (\law{D5}) and never appears at run
time: the adaptivity of (d) with the run-time determinism of (a),
at the scope of a project.}}
\label{fig:couplings}
\end{figure*}
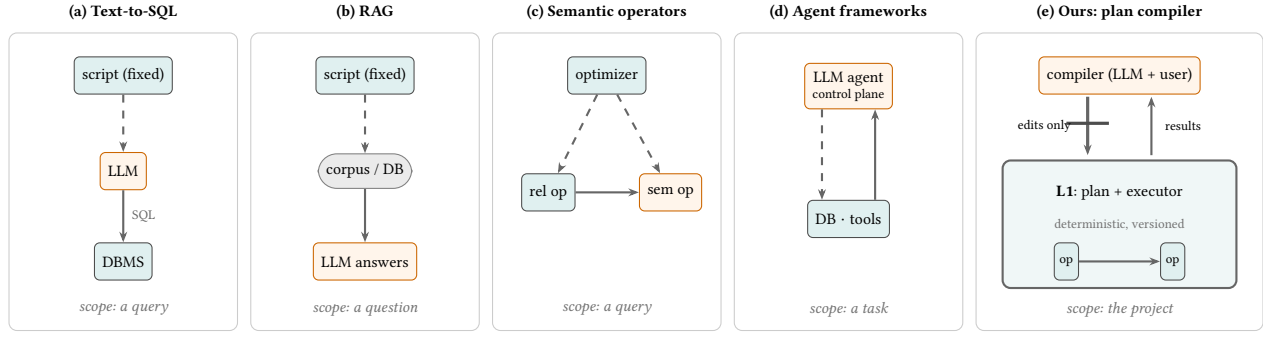

\begin{table}[t]
\centering
\footnotesize
\setlength{\tabcolsep}{3pt}
\caption{{LLM--database couplings: the LLM's seat, the
scope of one interaction, and what persists across interactions
(Figure~\ref{fig:couplings} shows who controls execution).}}
\label{tab:couplings}
\begin{tabular}{@{}l l l l@{}}
\toprule
\textbf{Coupling} & \textbf{LLM's seat} & \textbf{Scope} & \textbf{Persists} \\
\midrule
Text-to-SQL~\cite{katsogiannisNL2SQL} & One-shot compiler & A query & DB tables \\
RAG~\cite{lewisRAG} & Answerer & A question & Corpus \\
Semantic ops~\cite{patel2024lotus} & Data-path operator & A query & None \\
Agent frameworks~\cite{luAIScientist} & Control plane & A task & Transcript \\
\textbf{Ours} & Plan compiler & A project & Record \\
\bottomrule
\end{tabular}
\end{table}

{%
{%
\textbf{Where the LLM sits.} This seat is not the one existing
LLM--database couplings give the model, and the difference is
easiest to see side by side (Figure~\ref{fig:couplings},
Table~\ref{tab:couplings}; {\fullref{sec:positioning}} discusses
each in detail). Text-to-SQL and retrieval-augmented generation (RAG)
keep control deterministic but \emph{fixed}: a hardwired pipeline
is trustworthy precisely because it decides nothing, and for the
same reason cannot adapt when a translation misses or a retrieval
fails. Semantic operators adapt under a deterministic query
optimizer, the structure we advocate, but the stochastic call sits
on the \emph{data path}, invoked by the executor itself, and its
scope is a single query: nothing persists to reuse, maintain, or
audit afterwards. Agent frameworks adapt at the scope of a whole
task, but by handing the control plane to the stochastic component,
which then decides, executes, and grades itself. The two-stratum
design is the remaining combination: control that \emph{adapts},
because the compiler keeps revising a standing plan, yet is
\emph{deterministic at run time}, because the executor owns every
run-time decision and the LLM never appears in it, and that
persists at the scope of a project. Read left to right, prior
designs move the LLM ever closer to the database; we move the
database's discipline into the loop that runs the research.}}

{%
\textbf{What this buys.} Reliability and economy are the headline:
{the tagged guarantees\psecref{sec:functionality} follow from the
rules, and the classical optimization machinery then applies with
exact semantics\psecref{sec:performance}. Two further
properties fall out. The system becomes \emph{transparent}: the
plan is an explicit object the researcher can read, query, and
edit\psecref{sec:transparency}. And it becomes
\emph{collaborative}: the project is one small database instance,
so sharing it shares the research, down to peer review conducted
over the submitted instance\psecref{sec:collab}.}
{The four
properties share one objective: to minimize the manual work the
researcher must contribute. Versioning, re-running, verification, and hand-off are mechanical and belong to the system; forming hypotheses, interpreting results, and choosing the next experiment are the creative part of research and cannot be delegated. The design aims to move the researcher's time from the first kind of work to the second, so that her effort is spent on the ideas and the decisions that are genuinely important and intellectual.}}

{%
We make the following \textbf{contributions.}
\begin{enumerate}[leftmargin=1.4em]
\item The diagnosis and the \emph{confinement thesis}: a stochastic
NL$\to$DAG compiler over a deterministic, versioned executor, with
five design rules governing the
{boundary
between the two strata, the single interface through which compiler
output enters the executor and executed results flow back}
(Sections~\ref{sec:intro}--\ref{sec:model}).
\item The requirement-to-technique map (Table~\ref{tab:mapping}):
four properties and eighteen tagged requirements, each with its
database counterpart and its delivering mechanism, showing that
prior agent--data-systems work fills only the performance column
(Table~\ref{tab:desiderata}).
\item Each guarantee derived from the rules, walked
on a single running example
({\fullrefs{sec:functionality}{sec:collab}}).
\item A prototype with preliminary
results{\psecref{sec:prelim}} and %
{the open problems the design exposes}
{\ifarxiv(both in the full version, in preparation)\else(Section~\ref{sec:agenda})\fi}.
\end{enumerate}}

\section{What an ADRS Should Guarantee}
\label{sec:desiderata}

\subsection{Four properties: reliable, non-wasteful, transparent, collaborative}
\label{sec:properties}

{%
An ADRS should have four properties, each a bundle of tagged
requirements. Table~\ref{tab:mapping} explains each tag: name, the requirement it states, the database capability
it mirrors, an example from Section~\ref{sec:episode}, and our
mechanism that will enforce it. 
\textbf{Reliable (R)}: what the system reports is true of what it
computed. \rtag{1} \emph{versioning}: every artifact version is
kept and addressable; \rtag{2} \emph{idempotence}: the same
request over an unchanged project returns the same result;
\rtag{3} \emph{rollback}: failures and interruptions restore a
consistent state; \rtag{4} \emph{provenance}: every result names
the inputs that produced it; \rtag{5} \emph{evidence}: reported
numbers come only from executions; \rtag{6} \emph{isolation}:
concurrent work cannot silently overwrite.
\textbf{Non-wasteful (N)}: nothing is paid for twice. \ntag{1}
\emph{incrementality}: a change recomputes only what it
invalidates; \ntag{2} \emph{reuse}: unchanged sub-computations are
never redone; \ntag{3} \emph{scheduling}: work is ordered under
cost and attention budgets; \ntag{4} \emph{approximation}: cheap
trials precede expensive runs.
\textbf{Transparent (T)}: the researcher can see and direct the work
without reverse-engineering a transcript. \ttag{1}
\emph{legibility}: current and planned work is readable structure;
\ttag{2} \emph{introspection}: the state answers questions about
itself; \ttag{3} \emph{progress}: cost and completion estimates
are visible; \ttag{4} \emph{steerability}: pause, pin, and
reorder are obeyed commands.
\textbf{Collaborative (C)}: results, and the state behind them, are
usable by people who did not produce them.
{\ctag{1} \emph{onboarding}: a newcomer joins
without human re-explanation}; \ctag{2} \emph{sync}: changes reach
others as diffs with their impact; \ctag{3} \emph{concurrency}:
parallel contributions merge or conflict explicitly; \ctag{4}
\emph{reviewability}: a third party can verify any claim against
the executions behind it.
Section~\ref{sec:episode} shows what the absence of these
requirements costs in practice.}

\begin{table*}[t]
\centering
\footnotesize
\setlength{\tabcolsep}{4pt}
\caption{{The
eighteen requirements towards four properties (R/N/T/C). Each row gives the tag, its name, what it
requires, the database
capability it mirrors, the event of Section~\ref{sec:episode} that
shows its absence, and the mechanism that achieves it
({\ifarxiv Section~\ref{sec:model} and the full version\else Sections~\ref{sec:model}--\ref{sec:collab}\fi}).}}
\label{tab:mapping}
{%
\begin{tabular}{@{}l l p{0.24\textwidth} p{0.13\textwidth} p{0.23\textwidth} p{0.22\textwidth}@{}}
\toprule
\textbf{Tag} & \textbf{Name} & \textbf{Requirement} & \textbf{DBMS counterpart} & \textbf{Example (\S\ref{sec:episode})} & \textbf{Mechanism} \\
\midrule
\rtag{1} & Versioning     & every artifact version is kept and addressable            & MVCC, time travel          & six figures, no ``new plot''; the lost better setup   & append-only registry; references pin versions \\
\rtag{2} & Idempotence    & same request over an unchanged project, same result       & deterministic re-execution & repeated t-SNE differs; the fixed bug returns          & canonicalized-request and result caches; context from current versions \\
\rtag{3} & Rollback       & failure or interruption restores a consistent state       & WAL, savepoints, atomicity & NaN crash; half-applied two-file change                & append-only record; atomic edits; snapshot restore \\
\rtag{4} & Provenance     & every result names the exact inputs that produced it      & lineage                    & what is stale after the \fn{aug.py} change?            & per-execution read/write records \\
\rtag{5} & Evidence       & reported results come from executions only                & integrity constraints      & the asserted 94\%                                      & evidence gate: only evaluation outputs enter \\
\rtag{6} & Isolation      & concurrent work cannot silently overwrite                 & snapshot isolation, OCC    & two workers rewrite \fn{train.py}                      & versioned edits; conflicts detected at apply \\
\addlinespace
\ntag{1} & Incrementality & a change recomputes only what it invalidates              & incremental view maint.    & re-run all or skip too much; styling-only change       & field-level staleness propagation \\
\ntag{2} & Reuse          & unchanged sub-computations are never redone               & materialized views, work sharing & t-SNE projection recomputed for every figure     & content-hash memoization; operator decomposition \\
\ntag{3} & Scheduling     & work is ordered under cost and attention budgets          & cost-based optimization    & overnight runs; office-hour questions; idle servers    & recorded per-node costs drive schedule and ETA \\
\ntag{4} & Approximation  & cheap trials precede expensive runs                       & LIMIT, speculation         & five-minute subset run catches the bad learning rate   & small proxy nodes gate full runs \\
\addlinespace
\ttag{1} & Legibility     & current and planned work is readable structure, not prose & query plans, EXPLAIN       & rewording prompts to redirect the agent                & the plan is an explicit DAG the user can edit \\
\ttag{2} & Introspection  & the state answers questions about itself                  & catalog queries            & why is the comparison marked stale?                    & plan and record are queryable \\
\ttag{3} & Progress       & cost and completion estimates are visible                 & progress indication        & stuck or thinking? finished when?                      & per-node ETAs from recorded costs \\
\ttag{4} & Steerability   & pause, pin, and reorder are obeyed commands               & workload management        & declaring the baseline settled                         & freeze and reorder as plan operations \\
\addlinespace
\ctag{1} & Onboarding     & full project state transfers without narration            & replication                & the inherited chat log                                 & copy the instance; attach and query \\
\ctag{2} & Sync           & changes reach others as diffs with their impact           & change data capture        & updates travel by message and screenshot               & version diffs; staleness across replicas \\
\ctag{3} & Concurrency    & parallel contributions merge or conflict explicitly       & concurrency control        & simultaneous write-up and experiment edits             & optimistic control across replicas \\
\ctag{4} & Reviewability  & third parties verify claims against executions            & auditing                   & judging a paper from the PDF alone                     & read-only queries over the instance \\
\bottomrule
\end{tabular}}
\end{table*}

\subsection{{Running example: the augmentation study}}
\label{sec:episode}

{%
{We follow one project from its first request to its submission for review.} A researcher asks an
ADRS whether data augmentation improves a CIFAR-10 classifier by at
least one accuracy point; the agent sets up the obvious pipeline
(load the data, train a baseline and an augmented model, evaluate
both, compare). The events that follow are ordinary, and each
exposes a missing requirement, which we tag ({Table~\ref{tab:mapping}}).
}

{%
{Curious
whether the augmented model separates the classes more cleanly, she
asks the agent to visualize the model's input and output embeddings with 2D t-SNE.}
{A week later she asks for the t-SNE again. The agent does
not recognize the repeat: it {recomputes the embeddings, projects them into 2D space (the main bottleneck),} and rewrites the plotting code, paying the full cost again
(\ntag{2}), and since the rewritten script differs from the
original, so does the figure (\rtag{2}).}
{By now half a dozen figures have accumulated across
sessions, baseline and augmented, different styling passes, the repeated ask, and an instruction like ``use the new plot'' no longer has a
grounded referent (\rtag{1}).} The agent reports 94\%
accuracy, a number that appears in no execution log; the LLM
asserted it, and the comparison downstream consumed it (\rtag{5}).
{She then improves \fn{aug.py}. Which results are now stale? The augmented training run, its evaluation, comparison with baseline, {and output embeddings, but not the input embeddings and their 2D projections, if they were computed from a predefined embedding model or frozen during the training.} Telling these
apart requires knowing exactly what each result was derived from
(\rtag{4}).}
{Without that knowledge the agent either re-runs the
whole pipeline, burning GPU-hours on unaffected branches, or skips
too much and leaves a stale verdict standing}
(\ntag{1}), and a cosmetic change to plot
styling should not re-run the t-SNE projection at all (\ntag{1}, at finer grain).
The augmented run itself takes four GPU-hours; a five-minute trial
on a small subset would have exposed the mis-set learning rate
before the bill (\ntag{4}).
{One night a training run crashes at epoch 12, leaving half-written
checkpoints. Another day she interrupts the agent midway through a
change that spans two files: \fn{train.py} is already switched to
the new data split, the evaluation script still uses the old one.
Nothing crashes and everything still runs, but every accuracy
produced afterwards is quietly meaningless, because the model is
now evaluated on data it was trained on. Recovery has to restore a
coherent whole, not merely resume from the last checkpoint
(\rtag{3}).}
{A later augmentation variant turns out to \emph{lower}
accuracy, so she tries to return to last week's better setup by
hand: the code she can dig out of the chat, but the exact library
versions and data ordering are gone, the recovered state never
quite reproduces the old number, and no one can say with authority
which version actually scored better, short of her having
versioned every piece by hand
{(\rtag{1}, \rtag{3})}.}
{She parallelizes the work to save time: one agent worker tunes the learning-rate
schedule while another refactors the data loader. Both rewrite
\fn{train.py}, and one silently overwrites the other (\rtag{6}).}
{And once the chat outgrows the LLM's context window
and gets compacted, the agent abruptly forgets the styling and code
conventions it was told weeks earlier~\cite{autoresearchAI}
(\rtag{2} again).}
{Through all of this she spends evenings correcting work the agent
got subtly wrong, an overhead that measurement suggests can exceed
doing the task oneself~\cite{metrRCT}. Worse, corrections do not
stay fixed. She catches the evaluation normalizing test images
with statistics computed on the training set and has the agent fix
it; the accuracies and figures already derived from the wrong
preprocessing stay in the draft until she hunts each one down, and
a week later, after the chat is compacted, the agent rewrites the
preprocessing from memory and the bug is back. A correction is
done only when everything derived from the wrong version has been
found (\rtag{4}) and refreshed (\ntag{1}), and when the fix cannot be silently undone
(\rtag{2}).}}
{None of these pains is unsolvable by hand. She could version every
artifact herself, write a checker for every claim, rebuild
environments from notes, and proofread each step; diligent
researchers do exactly this, and it consumes the hours the agent
was meant to save. The question is how the guarantees can come from
the system itself, leaving the human in the loop only where her
judgment is the point.}

{%
{Steering, in fact, is its own struggle: what the agent intends to
do next exists nowhere she can inspect}, so
{to redirect the
agent she rewords her instruction until a phrasing happens to land,
and she cannot simply look at what the agent intends to do}
(\ttag{1}). She wants to see what is planned, running, and done,
and when it will finish, so that four-hour runs go overnight and interesting and important
questions wait for office hours (\ttag{3}, \ntag{3}); to ask why
the comparison is marked stale (\ttag{2}); %
{and to declare
the finished baseline settled, so that nothing the agent does next
can alter its results (\ttag{4})}.

{As the project grows beyond one person,} a collaborator
joins and needs the whole picture, which today means %
{inheriting a chat log in which decisions, dead
ends, and stale numbers all look alike, then asking her to narrate
the rest (\ctag{1}). When the two work in parallel,
she on the experiment and the collaborator on the write-up,
updates travel by message and screenshot, and each learns of the
other's changes late or not at all (\ctag{2}, \ctag{3}).}
{And when the study is finally written up and submitted, its reviewer faces the same problem in the extreme: deciding, from a PDF alone, whether the headline number really is the output of the pipeline
and can be reproduced (\ctag{4}).}}

\subsection{{Why agents get none of this for free}}
{%
{In a classical DBMS these guarantees are not features to be added;
they are consequences of the execution model. Operators are
deterministic functions over explicit state, so idempotence,
provenance, and exact caching hold automatically, and decades of
work went into delivering them fast. An ADRS inherits none of this:
its steps are stochastic, and what the LLM has concluded so far
is not explicit anywhere. The guarantees must be re-established by
architecture.} Table~\ref{tab:desiderata} scores existing systems
against the taxonomy, and its column pattern tells the story.
{Agent-first data systems~\cite{liuAgentFirst} and AI-analytics
runtimes~\cite{russoDeepResearch}} fill the non-waste rows,
optimizing agent workloads aggressively, while guaranteeing almost
nothing in the reliable rows. Durable workflow
engines~\cite{stonebrakerACDC} fill reliable rows, but under the
premise that failures are crashes; they assume each step is
semantically correct, the one assumption an ADRS cannot make.
Deterministic build systems (Make, Bazel,
Nextflow~\cite{nextflow,mokhovBuild}) supply exactly the executor
half of our proposal:
{they re-run precisely the
targets a source change affects, deterministically, but they run
fixed plans written by trusted humans, with no language front end,
no goal, and no notion of a claim needing evidence
(Section~\ref{sec:model} returns to them).}
{No existing column provides the reliable rows and
the non-wasteful rows together for a research process, and outside
ours, the transparent and collaborative rows are nearly empty.}}

\begin{table}[t]
\centering
\footnotesize
\setlength{\tabcolsep}{3pt}
\caption{
{Guarantees across systems. \yes\,=\,supported;
\pt\,=\,partial; \no\,=\,absent; \na\,=\,not applicable. Ours is a
vision: \yes\ means ``by design.''}}
\label{tab:desiderata}
\begin{tabular}{@{}l ccccccc@{}}
\toprule
 & \rh{Classical DBMS}
 & \rh{{Status-quo ADRS~\cite{luAIScientist,gottweisCoScientist}}}
 & \rh{{Build systems~\cite{nextflow,mokhovBuild}}}
 & \rh{{Agent-first system~\cite{liuAgentFirst}}}
 & \rh{{AI-analytics runtime~\cite{russoDeepResearch}}}
 & \rh{{AC/DC workflows~\cite{stonebrakerACDC}}}
 & \rh{\textbf{Ours}} \\
\midrule
Confines nondeterminism        & \na & \no & \na & \no & \no & \no & \yes \\
Research-process scope          & \no & \pt & \pt & \no & {\pt} & {\no} & \yes \\
\addlinespace
\multicolumn{8}{@{}l}{\textit{Reliable}} \\
\quad R1 Versioning             & \yes & \no & \pt & \pt & {\no} & \pt & \yes \\
\quad R2 Idempotence            & \yes & \no & \yes & \no & \no & \yes & \yes \\
\quad R3 Rollback / recovery    & \yes & \no & \pt & \pt & \no & \yes & \yes \\
\quad R4 Provenance             & \pt  & \no & \yes & \pt & {\no} & \yes & \yes \\
\quad R5 Evidence-backed results & \yes\textsuperscript{*} & \no & \na & \no & \pt & \pt & \yes \\
\quad R6 Isolation              & \yes & \no & \pt & \pt & \no & {\pt} & \yes \\
\addlinespace
\multicolumn{8}{@{}l}{\textit{Non-wasteful}} \\
\quad N1 Incremental maint.     & \yes & \no & {\yes} & \pt & \no & \no & \yes \\
\quad N2 Memo / work sharing    & \yes & \pt & \yes & \yes & \yes & \no & \yes \\
\quad N3 Cost-based scheduling  & \yes & \no & \no & \yes & \yes & \no & \yes \\
\quad N4 Approximation          & \pt  & \no & \no & \yes & \yes & \no & \yes \\
\addlinespace
\multicolumn{8}{@{}l}{\textit{Transparent}} \\
\quad T1 Legible plan           & \pt  & \no & \pt & \no & \pt & \pt & \yes \\
\quad T2 Introspection          & \yes & \no & \pt & \pt & \no & \pt & \yes \\
\quad T3 Progress / ETA         & \pt  & \no & \pt & \no & \pt & \no & \yes \\
\quad T4 Steerability           & \no  & \pt & \no & \pt & {\no} & \pt & \yes \\
\addlinespace
\multicolumn{8}{@{}l}{\textit{Collaborative}} \\
\quad C1 Onboarding             & {\pt} & \no & \pt & \no & \no & \no & \yes \\
\quad C2 Sync / change feed     & \pt & \no & \no & \no & \no & \no & \yes \\
\quad C3 Concurrent merge       & \yes & \no & \pt & \pt & \no & \pt & \yes \\
\quad C4 Reviewability          & {\pt} & \no & \pt & \no & \no & \no & \yes \\
\bottomrule
\end{tabular}
\\[2pt]
{\footnotesize \textsuperscript{*}for data integrity, not research intent.}
\end{table}

\section{The Two-Stratum Model}
\label{sec:model}

{%
The design separates a research project into two strata.
\textbf{L1} holds
{the project's record: the plan that the user and the agent
have agreed to operate on, everything computed under that plan, and
the full execution history. Agreement is the consistency criterion
for this state: every change moves it from one plan both parties
can see to another.} All of L1 is deterministic and versioned.
\textbf{L2} decides what to compute \emph{next}: it is where the
LLM and the user live, and it touches L1 only by editing the plan
and reading results. Figure~\ref{fig:arch} shows both strata on the
running example: 
{the compiler's components} across the top; below
them, the augmentation pipeline as L1's graph, with artifacts drawn
as pills and operators as boxes; at the bottom, the three stores
that persist it all; and between the strata, the two arrows that
are the only traffic across the boundary, edits going down and
executed results coming up.
{We specify the design as a small
type system (Figure~\ref{fig:types}) rather than as an architecture
diagram alone, because the discipline we are claiming is easiest to
verify as types: every name in that figure denotes a function,
except the single stochastic arrow, \fn{edit}.}
{%
Figure~\ref{fig:types} states the design as sixteen declarations,
(S1)--(S16), each annotated with its role; every name in it also
appears in Figure~\ref{fig:arch}, so the two figures describe one
system at two zoom levels. The rest of this section walks them.
Section~\ref{sec:l1} covers the deterministic stratum,
(S1)--(S9); Section~\ref{sec:l2} covers the compiler,
(S10)--(S16), and traces one request of the running example through
both strata; Section~\ref{sec:ctx} examines one declaration,
\fn{view} (S11), more closely, because how the compiler's context is
built is where this design departs most from today's agents;
Section~\ref{sec:rules} then states the five design rules,
\law{D1}--\law{D5}, that constrain the declarations and from which
the guarantees follow; and Section~\ref{sec:buildsys} asks what
separates L1 from an ordinary build system.}}

\begin{figure*}[t]
\centering
\begin{tikzpicture}[x=1pt,y=1pt,
  every node/.style={font=\footnotesize},
  op/.style={draw=black!70, fill=teal!12, rounded corners=2pt,
             minimum height=16pt, inner sep=3pt},
  art/.style={draw=black!50, fill=black!8, rounded corners=7pt,
              minimum height=14pt, inner sep=3pt},
  st/.style={draw=violet!70!black, fill=violet!8, rounded corners=2pt,
             minimum height=16pt, inner sep=4pt},
  l2c/.style={draw=orange!80!black, fill=orange!8, rounded corners=2pt,
              minimum height=16pt, inner sep=4pt},
  arr/.style={-{Stealth[length=5pt]}, black!60, thick}]

\node[draw=orange!80!black, thick, rounded corners=4pt, fill=orange!4,
      minimum width=470pt, minimum height=64pt] (L2) at (240,150) {};
\node[anchor=north west, font=\footnotesize\bfseries,
      text=orange!60!black] at (8,178) {L2 --- the compiler (S10--S16): stochastic, turns NL requests into plan edits; LLM + user};

\node[l2c] (req)  at (60,145)  {\shortstack{Request\\\scriptsize S10 $\cdot$ NL ask}};
\node[l2c, thick] (edit) at (150,145) {\shortstack{\textbf{Compiler (LLM)}\\\scriptsize S12 $\cdot$ the only \textsf{Stoch}}};
\node[l2c] (delta) at (247,145) {\shortstack{DagEdit $\delta$\\\scriptsize S13 $\cdot$ add $\cdot$ put $\cdot$ freeze}};
\node[l2c] (ec)   at (342,145) {\shortstack{EditCache\\\scriptsize S14 $\cdot$ $H(\mathrm{canon},\mathrm{digest})$}};
\node[l2c] (bel)  at (438,145) {\shortstack{State $s$\\\scriptsize S15 $\cdot$ goal $\cdot$ Pred $\cdot$ notes}};

\draw[arr] (req) -- (edit);
\draw[arr] (edit) -- (delta);
\draw[arr] (edit.north) .. controls (200,166) and (290,166) .. (ec.north);
\draw[arr] (bel.south) -- ++(0,-9) -| (edit.south);

\draw[arr, dashed] (105,86) -- node[left, align=right, font=\scriptsize]
  {\texttt{view} (S11):\\ assembled context} (105,122);
\draw[arr, very thick] (247,122) -- node[left, align=right, font=\scriptsize]
  {\texttt{apply}: \textbf{edits only};\\ LLM output enters as a\\ versioned Artifact, then \emph{runs}} (247,86);
\draw[arr, very thick] (438,86) -- node[right, align=left, font=\scriptsize]
  {\texttt{observe} (S16): \textbf{executed}\\ \textbf{results only} (rule D5)} (438,122);

\node[draw=black!60, thick, rounded corners=4pt, fill=blue!3,
      minimum width=470pt, minimum height=120pt] (L1) at (240,20) {};
\node[anchor=north west, font=\footnotesize\bfseries, text=black!70]
  at (8,76) {L1 --- the record and its executor (S1--S9): \texttt{run} (S8) is a \emph{function}; never calls an LLM};

\node[art] (aug)   at (48,48)  {\shortstack{\texttt{aug.py}\\\scriptsize Artifact (S1)}};
\node[art] (data)  at (48,22)  {\texttt{cifar.npz}};
\node[op]  (train) at (122,35) {\shortstack{\texttt{train\_aug}\\\scriptsize Operator (S3) $\cdot$ Exec}};
\node[art] (model) at (196,35) {\texttt{model}};
\node[op]  (eval)  at (258,35) {\shortstack{\texttt{eval}\\\scriptsize Eval}};
\node[art] (acc)   at (318,35) {\texttt{acc\_aug}};
\node[op]  (cmp)   at (384,35) {\shortstack{\texttt{compare}\\\scriptsize Eval}};
\node[art] (verd)  at (450,35) {\texttt{verdict}};
\node[op]  (tsne)  at (258,8)  {\shortstack{\texttt{tsne}\\\scriptsize Exec}};
\node[art] (plot)  at (330,8)  {\texttt{tsne.png}};

\draw[arr] (aug) -- (train);
\draw[arr] (data) -- (train);
\draw[arr] (train) -- (model);
\draw[arr] (model) -- (eval);
\draw[arr] (eval) -- (acc);
\draw[arr] (acc) -- (cmp);
\draw[arr] (cmp) -- (verd);
\draw[arr] (model) |- (tsne);
\draw[arr] (tsne) -- (plot);

\node[st, font=\scriptsize] (reg) at (88,-25)  {Registry $R$ (S5): append-only artifacts};
\node[st, font=\scriptsize] (log) at (237,-25) {ExecLog $E$ (S6): status $\cdot$ cost $\cdot$ reads};
\node[st, font=\scriptsize] (cch) at (402,-25) {Cache $K$ (S7): sig $\mapsto$ ref $\cdot$ IVM via affects (S9)};

\end{tikzpicture}
\caption{
{The two strata on the running example, with each
component labeled by its declaration (S1)--(S16) in
Figure~\ref{fig:types}. All nondeterminism lives in the compiler
(top); L1 (bottom) is a deterministic, versioned dataflow engine
over immutable artifacts (pills, S1) and operators (boxes, S3).
Three paths cross the boundary: \texttt{view} (S11) carries context
up, \texttt{apply} carries edits (S13) down, and \texttt{observe} (S16)
returns executed results only (rule \law{D5}).}}
\label{fig:arch}
\end{figure*}
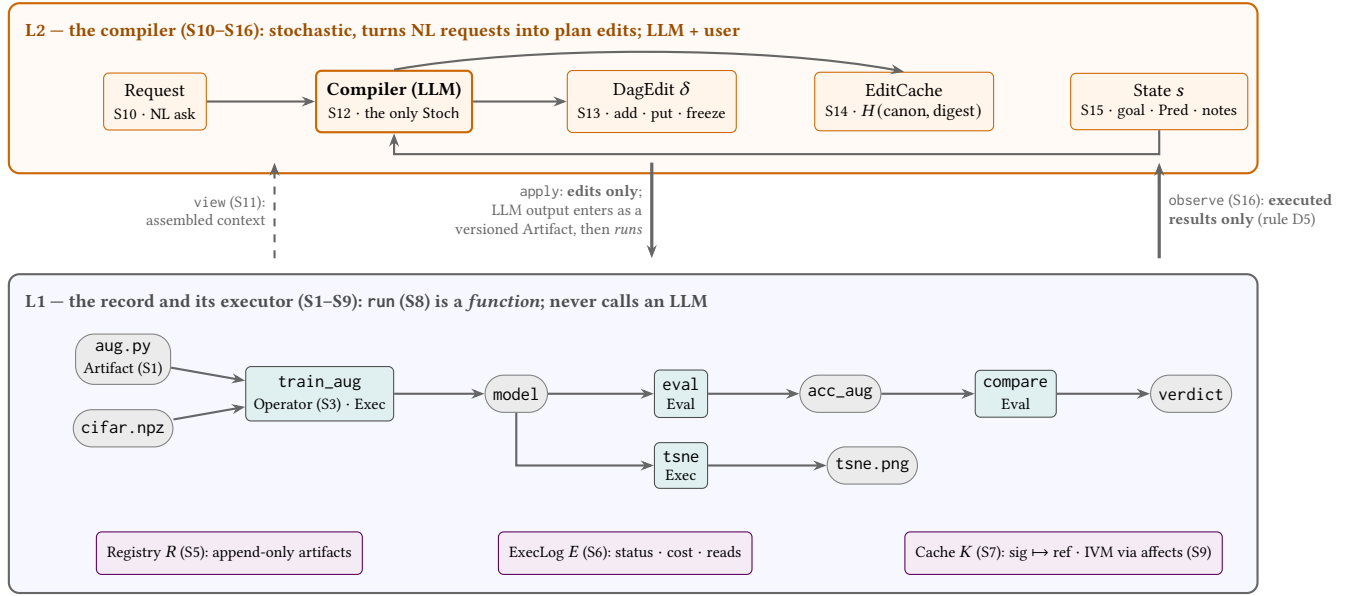

\begin{figure}[t]
\centering
\begin{lstlisting}[basicstyle=\ttfamily\footnotesize,
  keywordstyle=\bfseries, columns=fullflexible, frame=single,
  framerule=0.4pt, xleftmargin=2pt, xrightmargin=2pt]
== L1: the record and its executor (Fig.1, bottom)
S1  Artifact = {key, ver, value, deps}
    -- one immutable result; code and
    -- environments are artifacts too
S2  ver = (seq, time, contentHash)
    -- "the new plot" = a key's highest seq
S3  Operator = {reads: Set Key, writes: Key, rt}
    -- one step; train_aug reads aug.py and
    -- cifar.npz, writes model
S4  rt in {Exec, Eval, Diff, ...}
    -- fixed set of runtimes, all deterministic
S5  Registry R : Ref -> Artifact
    -- append-only store of all artifacts
S6  ExecLog E : (node, ver) -> ExecRec
    -- what a firing read/wrote/cost; status
S7  sig(node) = hash(rt, pinned input versions)
    Cache K : sig -> Ref
    -- same signature, same stored result
S8  run(Operator, R) -> Value    -- A FUNCTION
S9  a write flags its readers Stale iff
    affects(part read, delta)
    -- incremental maintenance, field-level

== L2: the compiler (Fig.1, top)
S10 Request = one NL ask, from user or agent
S11 view(Request, Snapshot) -> View
    View = closure as ArtifactRef | Digest
    -- Snapshot: the record at one instant;
    -- deterministic context assembly (Sec 3.3)
S12 edit(Request, State, View) ~> DagEdit, State
    -- the LLM; THE ONLY STOCHASTIC STEP
S13 DagEdit = add/rewire/delete node
    | put artifact | set goal | freeze subtree
S14 EditCache: hash(canon(Req), digest(View))
              -> DagEdit
    -- canon normalizes the request text;
    -- digest hashes the view; a repeated ask
    -- over an unchanged view replays the edit
S15 State = {goal, success: Pred, notes, budget}
    -- compiler's working memory; Pred is an Eval
S16 observe(State, View) -> State
    -- updated from executed results only
\end{lstlisting}
\caption{
{The design as sixteen declarations. (S1)--(S9) form the deterministic stratum of
Figure~\ref{fig:arch} (bottom): the data, the steps, the three
stores, and the executor. (S10)--(S16) form the compiler
(Figure~\ref{fig:arch}, top); \fn{edit} (S12) is the only
stochastic step, and \fn{view} (S11) and \fn{observe} (S16) are
the only ways context leaves L1 and results return.
Sections~\ref{sec:l1}--\ref{sec:ctx} walk the lines in order.}}
\label{fig:types}
\end{figure}

{%
\subsection{{L1: the project as a query plan}}
\label{sec:l1}
The unit of storage is the \ty{Artifact} (S1): an immutable value
(a dataset, a trained model, a figure, and, importantly, code and
environments too) identified by a key and a version, where the
version (S2) records a content hash and the exact input versions.
{The unit of computation is the \ty{Operator} (S3): a node that
names the artifact keys it reads, the single key it writes, and a
runtime drawn from a fixed deterministic set (S4). In Figure~\ref{fig:arch}, \fn{train\_aug} is an
operator that reads \fn{aug.py} and \fn{cifar.npz} under the
\ty{Exec} runtime, which runs code in a pinned environment, and
writes \fn{model}; \fn{compare} reads the two accuracies under the
\ty{Eval} runtime, which computes a check, and writes
\fn{verdict}. An operator touches nothing beyond its declared
ports.} To fire a node,
{L1's \emph{executor}} pins
each 
{input key} to
its current version and hashes the combination into a signature
(S7); a hit in the cache \ty{K} means the result already exists,
otherwise the node executes (\fn{run}, S8, a function of its
pinned inputs) and the new artifact is appended. The append-only
registry of artifacts (S5) and the execution log (S6) together form
what we will call the \emph{record}. Nothing in it is overwritten, and
every execution appends what it read, wrote, and
{cost}.
{Because operators read only their declared ports and
write only their output, dependency has a fixed meaning: node $m$
depends on node $n$ exactly when $m$ reads the key $n$ writes. When
a write gives a key a new version, the executor flags that key's
readers \ty{Stale} (S9),} but only if the part they read actually
changed: 
{in the running
example, \fn{compare} reads only the accuracy field of each
evaluation output, so a change that leaves both accuracies intact
does not touch it.} This is incremental view maintenance at field
granularity, applied to a research pipeline (\ntag{1}).}

{%
\subsection{{L2: the LLM as plan compiler}}
\label{sec:l2}
The compiler receives a \ty{Request} (S10), from the user or from
the agent's own planning, and sees a \ty{View} (S11):
{a read-only projection of the current
\emph{snapshot},
{an immutable read of the record as it stands at one instant}. Concretely,
a view is a value listing the plan's nodes and edges, each node's status and current version, and the goal; how this value is
assembled, and why that matters, is the subject of
Section~\ref{sec:ctx}.}
{It responds with a \ty{DagEdit} (S13): add or rewire nodes, put a
new artifact (say, code it just wrote), revise the goal, freeze a
subtree. Producing this edit is \fn{edit} (S12), the only
stochastic step in the system; applying the edit and
re-materializing the graph are deterministic.}
{Because \fn{edit} is stochastic, calling it
twice on the same question could yield two different plans, so the
compiler never calls it twice on the same question. This is the job
of the \ty{EditCache} (S14): before the LLM is invoked, the
request's text is normalized by \fn{canon}, so that incidental
rephrasing does not defeat matching, and paired with a hash of the
view, \fn{digest}; if this pair has been seen before, the recorded
edit is replayed and the LLM is not consulted at all.} The compiler also keeps a small working \ty{State} (S15): the goal,
the success criterion, the budget, and its \emph{notes}, free-form
working annotations such as hypotheses and decisions, written by
the LLM or by the user. Notes live in L2 as commentary and are
never treated as results. The state is updated by \fn{observe}
(S16) from executed L1 results, never from the compiler's own
assertions.}

\begin{figure*}[t]
\centering
\begin{tikzpicture}[x=1pt,y=1pt,
  every node/.style={font=\scriptsize},
  lab/.style={anchor=west, font=\scriptsize\bfseries, text=black!80},
  tk/.style={draw=black!60, fill=black!6, minimum width=24pt, minimum height=13pt, inner sep=1pt},
  sum/.style={draw=black!50, dashed, fill=black!14, rounded corners=3pt, minimum height=13pt, inner sep=2pt},
  llm/.style={draw=orange!80!black, fill=orange!8, rounded corners=2pt, minimum height=13pt, inner sep=3pt},
  card/.style={draw=black!45, fill=yellow!20, minimum width=24pt, minimum height=11pt, inner sep=1pt},
  onn/.style={draw=teal!70!black, thick, fill=teal!16, rounded corners=2pt, minimum height=13pt, inner sep=2pt},
  offn/.style={draw=black!40, fill=black!5, rounded corners=2pt, minimum height=13pt, inner sep=2pt, text=black!55},
  note/.style={anchor=west, text=black!55, font=\scriptsize\itshape},
  arr/.style={-{Stealth[length=4pt]}, black!60},
  darr/.style={-{Stealth[length=4pt]}, black!60, dashed}]

\node[lab] at (0,168) {(a) Sequential accumulation};
\node[lab, font=\scriptsize] at (0,157) {\textmd{(ReAct-style chat agents)}};
\foreach \i in {1,...,7} { \node[tk] at (150+\i*28,162) {$t_{\i}$}; }
\node at (388,162) {$\cdots$};
\node[note] at (390,162) {cost grows with history};

\node[lab] at (0,128) {(b) Lossy compaction};
\node[lab, font=\scriptsize] at (0,117) {\textmd{(Cursor, Claude Code)}};
\node[sum, minimum width=76pt] at (206,122) {summary$(t_1\ldots t_5)$};
\node[tk] at (262,122) {$t_{6}$};
\node[tk] at (290,122) {$t_{7}$};
\node[tk] at (318,122) {$t_{8}$};
\node at (346,122) {$\cdots$};
\node[note] at (390,127) {irreversible; details vanish,};
\node[note] at (390,116) {outputs drift};

\node[lab] at (0,86) {(c) External memory};
\node[lab, font=\scriptsize] at (0,75) {\textmd{(MemGPT, memory banks)}};
\node[llm] (cllm) at (170,80) {LLM};
\node[draw=black!40, rounded corners=3pt, minimum width=140pt, minimum height=34pt] (cstore) at (285,79) {};
\node[card] at (238,86) {note};
\node[card] at (266,86) {note};
\node[card] at (294,86) {note};
\node[card] at (322,86) {note};
\node[card] at (252,72) {note};
\node[card] at (308,72) {note};
\draw[arr] (cllm.north) to[bend left=18] node[above, font=\tiny]{writes own prose} (cstore.north west);
\draw[darr] (cstore.south west) to[bend left=14] node[below, font=\tiny]{top-$k$ by similarity} (cllm.south);
\node[note] at (390,91) {unverified prose in;};
\node[note] at (390,80) {similarity-ranked recall out;};
\node[note] at (390,69) {no versions, no lineage};

\node[lab] at (0,40) {(d) Dependency-closed};
\node[lab] at (0,29) {\phantom{(d) }assembly (ours)};
\node[onn] (dg) at (160,34) {goal};
\node[onn] (da) at (205,34) {\fn{aug.py}@v3};
\node[onn] (db) at (262,34) {\fn{model}@v7};
\node[onn, very thick] (dx) at (318,34) {\textbf{target}};
\node[offn] (dc) at (233,10) {\fn{eda.ipynb} $\to$ digest};
\node[offn] (dd) at (318,10) {\fn{baseline} $\to$ digest};
\draw[arr] (dg) -- (da); \draw[arr] (da) -- (db); \draw[arr] (db) -- (dx);
\draw[arr, black!35] (dc) -- (db); \draw[arr, black!35] (dd) -- (dx);
\node[draw=violet!70!black, fill=violet!6, rounded corners=2pt, inner sep=3pt] (dview) at (410,34) {\ty{View}};
\draw[arr] (dx.east) -- (dview.west);
\node[note] at (150,-10) {on-path = full fidelity, off-path = digest; version-pinned, re-expandable from the registry};
\end{tikzpicture}
\caption{{How each family manages the LLM's context. (a)
Chat agents re-send a transcript that only grows. (b) On overflow
the prefix is compacted into prose, irreversibly. (c) Memory
systems store the LLM's own notes and retrieve them by similarity;
admission and recall are both
unverified{\psecref{sec:positioning}}. (d) We assemble context per
request from the dependency closure: the path to the target in
full, off-path ancestors as digests, every element version-pinned
and re-expandable.}}
\label{fig:ctxtax}
\end{figure*}
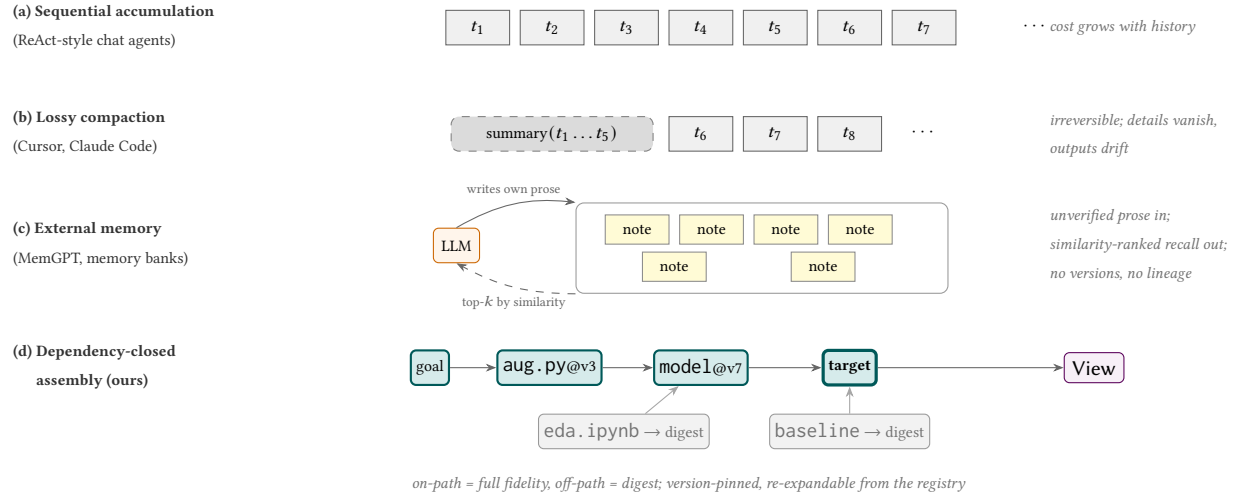

{%
To make this concrete, trace the episode's second visualization
request: ``plot the t-SNE again, marking the k-means clusters.''
The compiler canonicalizes the request and hashes it together with a
digest of the view; a verbatim repeat hits the \ty{EditCache} and
replays the previous edit
{without
invoking the LLM at all} (\rtag{2}). Otherwise the LLM proposes an
edit.
{What
happens next deserves care, because the first request left nothing
to reuse: it produced a single \fn{tsne} node that computes the
2D projection and renders the figure in one step, and no one
stored the projection. What the record does hold is the node's
cost profile, and it is telling: the projection took hours due to large data size, the
rendering a moment, and the projection is a few megabytes. Cost
statistics are exactly what a database optimizer feeds on, and
{the executor} uses them the same way. Seeing a second request over the
same prefix, it splits the coarse node into two sub-nodes,
\fn{tsne\_embed} and \fn{tsne\_plot}, and materializes the
expensive, compact intermediate for reuse; where the split of the
code cannot be derived mechanically, %
{the compiler (\fn{edit}, S12)}
is asked to propose it, and
{design rule \law{D1}
(determinism of L1, stated with the other rules in
Section~\ref{sec:rules}) makes the rewrite cheap to verify: re-run
on the recorded inputs and compare outputs byte-for-byte.} No one asked for the materialization. The repeat
revealed the projection as a shared prefix, the recorded costs
flagged it as the bottleneck, and later requests collect the
dividend: a styling change re-fires only \fn{tsne\_plot}, and
further t-SNE requests reuse the stored projection outright
(\ntag{2}, Table~\ref{tab:mapping}).} 
The k-means overlay itself is
{a new operator downstream of \fn{tsne\_embed}: a new
node, because it is a new computation. Re-styling the existing
figure would instead produce a new \emph{version} of
\fn{tsne\_plot}'s output.} 
{The executor applies the edit and materializes: the
two new nodes, \fn{kmeans} and the extended plot, run, while
\fn{tsne\_embed}'s stored projection is served from the cache,
because its signature (S7), computed from its runtime and inputs,
is unchanged. One might expect the split to complicate caching,
with one key for the whole t-SNE step and further keys per
sub-node. No such hierarchy is needed, because the graph already
encodes it: a request for the full figure resolves to
\fn{tsne\_plot}, whose signature pins \fn{tsne\_embed}'s output
transitively; a request needing only the projection resolves to
\fn{tsne\_embed} directly; and if \fn{tsne\_embed} is ever
updated, staleness reaches exactly its readers through the
dependency edge (S9), leaving unrelated sub-nodes untouched.} The figure is appended as
a new version, so ``the new plot'' has a referent (\rtag{1}), and
\fn{observe} updates the
{compiler's notes (Section~\ref{sec:l2})} from
the executed outputs.
{Throughout the exchange, nothing the LLM merely said
entered the record; only artifacts produced by executions did.}}

{%
{We emphasize two design choices.} First, the
\emph{goal itself is an artifact}: a versioned statement paired
with a deterministic success predicate (``\fn{compare} $\geq$
0.01''), so research intent is under version control and ``are we
done?'' is a computation, not an opinion. When the goal is too
exploratory to admit a crisp predicate on day one, the predicate is
provisional and its refinements are ordinary versioned edits
({\ifarxiv full version\else Section~\ref{sec:agenda}\fi, OP6}). Second, the \emph{evidence gate}:
{no claim about results, numeric or otherwise, can enter the
record except} as the output of a deterministic evaluation node,
{that is,
of an \ty{Operator} whose runtime is \ty{Eval}}. The LLM saying
``94\%'' produces, at most, an edit that adds such a node,
{and the accuracy that enters the record is whatever that node computes, not what the LLM predicted} (\rtag{5}).}

{%
{So far five tags have appeared: \rtag{1}, \rtag{2},
\rtag{5}, \ntag{1}, and \ntag{2}, cited where the declaration
that delivers them was introduced. The other thirteen are not
missing; they simply need more than one declaration. Neither
stratum owns a group of tags, and
{all eighteen are
treated in turn in \fullrefs{sec:functionality}{sec:collab}}.}}

{%
\subsection{Context from the graph, not the transcript}
\label{sec:ctx}}
{%
The type of \fn{edit} makes the LLM's entire input explicit:
{the request (S10), its
working state (S15), and the \ty{View} (S11)}.
Section~\ref{sec:l2} said what a view shows;
{here we specify how it is built. The \fn{view} function is
deterministic: it takes the nodes a request touches, walks their
\emph{dependency closure}, meaning those nodes together with all
their ancestors in the graph, and 
{assembles the view from that closure: the
plan slice Section~\ref{sec:l2} listed, plus the closure's
artifacts, each pinned to a version}.} The view is not ``the
conversation so far,'' and the difference matters more than it may
appear. Today's coding and research
agents keep one linear context that only grows; on overflow it is
\emph{compacted} into a lossy summary, and past that point the
agent forgets the plot style, the code conventions, the writing
voice it was following, because those details lived only in the
discarded prose.\footnote{
{A recent survey names evidence
preservation and provenance tracking among the persistent
challenges of research automation~\cite{autoresearchAI}; our
design supplies the mechanism.}} Here the details are versioned artifacts
and %
{notes
(Section~\ref{sec:l2})} on the graph: the plotting node's styling
decisions sit on the plotting node, and any request touching a
dependent gets them back at full fidelity, however old they are.
{Because the context of each request is
rebuilt from the graph, its size tracks the closure that is
actually relevant, not the length of the conversation; nothing is
ever evicted to make room, so there is no cliff to fall off.}
Nor must the
{closure} be fetched uniformly; each \ty{View}
element takes one of the two forms S11 declares: a full
{\ty{ArtifactRef}, a pinned
(key, version) reference whose value is included outright (both
forms are now declared in S11),} or a
{\ty{Digest}, a short deterministic summary of
an artifact %
{(for a table, its schema and row count;
for a training log, its final metrics; for a long document, its
opening section)} computed by an
ordinary L1 operator and therefore versioned like everything
else}. A natural policy fetches the dependency path of
the target in full while representing off-path ancestors by
digests. Context is thus compressed \emph{non-sequentially}, along
the graph's structure rather than the
{conversation's} order,
{and the compression is reversible: a
digest carries the (key, version) of the artifact it summarizes,
so re-expanding it is a registry lookup, and nothing about the
original is lost by sending the digest instead.}
Figure~\ref{fig:ctxtax} draws the contrast one family per row:
(a) chat agents re-send a transcript that only grows; (b) compaction
crushes its prefix irreversibly; (c) memory systems store the
LLM's own notes and recall them by
similarity{\psecref{sec:positioning}};
{(d) we assemble the closure. The design does not
need to remember well, because nothing is ever forgotten: whatever
matters is either in the record or exactly recomputable from it.}
Choosing {between full artifact and digest} per
{closure element} is a real optimization
question, but a \emph{scoped} one,
{ranging over one request's closure rather than over
everything ever said}{\ifarxiv\else{} (OP4)\fi}.
{Finally,
this construction is what makes the \ty{EditCache} of
Section~\ref{sec:l2} correct: the view is a deterministic function
of versioned state, so two identical requests over an unchanged
project see byte-identical views, and ``same request, same view''
becomes a sound cache key
{(\rtag{2})}.}}

\begin{figure}[t]
\centering
\begin{tikzpicture}[x=1pt,y=1pt,
  every node/.style={font=\scriptsize, align=center},
  sto/.style={draw=orange!80!black, fill=orange!8, rounded corners=2pt,
              minimum height=14pt, inner sep=3pt},
  sidenote/.style={anchor=east, font=\tiny, text=black!55, align=right},
  arr/.style={-{Stealth[length=4pt]}, black!60, thick}]

\draw[draw=orange!80!black, fill=orange!8, rounded corners=2pt]
  (52,84) rectangle (198,112);
\node at (125,102) {model weights};
\node[font=\tiny, text=black!55] at (125,91) {lossy $\cdot$ frozen $\cdot$ no lineage};

\draw[draw=black!50, fill=black!6, rounded corners=2pt]
  (52,44) rectangle (198,72);
\node at (125,62) {context window};
\node[font=\tiny, text=black!55] at (125,51) {per-call working set (buffer pool)};

\draw[draw=violet!70!black, fill=violet!8, rounded corners=2pt]
  (52,4) rectangle (198,32);
\node at (125,22) {the record: $R$ (S5) $+$ $E$ (S6)};
\node[font=\tiny, text=black!55] at (125,11) {versioned $\cdot$ provenanced base store};

\node[sidenote] at (50,98) {refresh $=$\\[-2pt] retrain};
\node[sidenote] at (50,58) {transient};
\node[sidenote] at (50,18) {persistent};

\draw[arr] (70,32) -- (70,44);
\node[font=\tiny, anchor=west] at (74,38) {\fn{view} (S11): deterministic fill};

\node[sto, font=\tiny] (ed) at (224,58) {\fn{edit}\\(S12)};
\draw[arr] (198,98) to[bend left=20] (ed.north);
\draw[arr] (198,58) -- (ed.west);

\draw[arr] (224,46) |- (200,18);
\draw[black!70, very thick] (212,11) -- (212,25);
\node[font=\tiny] at (212,30) {(\law{D5})};

\end{tikzpicture}
\caption{{The three storage tiers of an LLM-based system,
in database terms. The weights are a lossy, frozen materialized
view of the training corpus; the context window is a per-call
buffer pool; the record is the versioned base store. \fn{edit}
reads the two upper tiers, prior and evidence; its output reaches
the record only through the gate (\law{D5}). The families of
Figure~\ref{fig:ctxtax} are buffer-management policies for the
middle tier; \fn{view} is the DBMS-managed one.}}
\label{fig:tiers}
\end{figure}
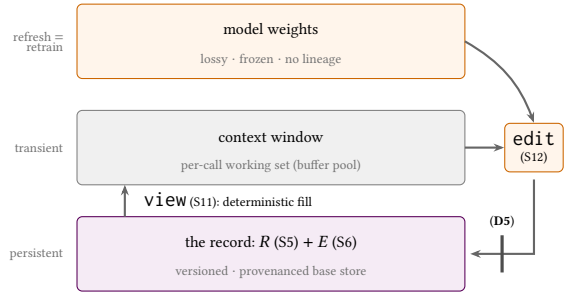

{%
{%
There is also a storage reading of this design choice. An LLM-based
system holds state in three tiers (Figure~\ref{fig:tiers}). The
model's \emph{weights} are, in database terms, a materialized view
of everything the model was trained on: lossy, frozen between
retrainings, with no per-fact addressability, no delete, and no
lineage. The \emph{context window} is a transient, per-call working
set, the buffer pool. And the \emph{record} is the versioned,
provenanced base store, the tier today's agents lack. Seen this
way, the families of Figure~\ref{fig:ctxtax} are buffer-management
policies for the middle tier: (a) admits everything and never
evicts, so the working set grows without bound; (b) evicts by
irreversible compaction; (c) is application-managed buffering, the
LLM itself deciding admissions in its own prose and recalling by
similarity. Ours is the DBMS-managed policy: \fn{view} fills the
working set deterministically from the base store, the staleness
rule (S9) keeps buffered copies coherent with the record, and
eviction costs nothing because every element is version-pinned and
re-fetchable. The tier the system cannot manage, the weights, is
also the tier it never needs to trust: whatever the project relies
on lives in the record, and a claim originating in the weights
enters it only through the gate of \law{D5}. In Codd's terms, the
design restores \emph{data independence}~\cite{codd1970}: the
project's knowledge is not entangled in anyone's weights, or
anyone's prose.}}

{%
\subsection{
{The design
rules behind the guarantees}}
\label{sec:rules}
{The guarantees of Section~\ref{sec:desiderata} must
not depend on the LLM behaving well, so we pin them to five rules
about how the strata may interact. Each later section points back
at the rule it relies on.}
\begin{description}[leftmargin=1.6em, itemsep=1pt]
\item[\law{D1} (determinism)] Every L1 runtime is a function of its
pinned inputs; re-materializing the graph reproduces byte-identical
artifacts.
\item[\law{D2} (append-only)] The {record
(registry and log, Section~\ref{sec:l1})} never overwrites; every
(key, version) pair is permanent.
\item[\law{D3} (confinement)] \fn{edit} is the only stochastic
relation in the system.
\item[\law{D4} (idempotence)] Answers served from
{either cache (\ty{K}, S7;
\ty{EditCache}, S14)} are
functions of their signature: the same request over the same view
yields the same edit, hence the same graph, hence (by \law{D1})
the same result.
{\item[\law{D5}
(evidence gate)] 
{The record accepts appends from \fn{run} (S8) alone; the compiler has no write path into it. An LLM claim
such as ``accuracy is 94\%'' can become, at most, a note in L2 or
an edit that adds an \ty{Eval} node, and what enters the record is
that node's executed output. In the other direction, the compiler's
state is updated by \fn{observe} (S16) reading executed L1
results, never by its own assertions. Either way, a claim becomes
part of the record only by running.}}
\end{description}
{Read the rules as a division of
labor. \law{D1} and \law{D2} make the record replayable and
traceable: any artifact can be reproduced by re-materialization,
and any result traced to the exact inputs that made it, which is
what versioning (\rtag{1}), rollback (\rtag{3}), provenance
(\rtag{4}), and exact reuse (\ntag{1}, \ntag{2}) stand on.
\law{D4} makes repeated requests reproducible (\rtag{2}), and
\law{D5} keeps unexecuted claims out of the record (\rtag{5}).}}

{%
\subsection{%
{Isn't L1
just a build system?}}
\label{sec:buildsys}
{Much of
it is, deliberately. A build system such as Make tracks
which targets a changed source file affects and rebuilds exactly
those,
{which
is the staleness rule (S9) in miniature};} Bazel and Nextflow~\cite{nextflow,mokhovBuild}
prove that deterministic, content-addressed DAG execution is
practical at scale, and we claim no novelty for L1's parts in
isolation. 
{What no build system has is the boundary, and every
construct of Figure~\ref{fig:types} that a build system lacks sits
on it: an untrusted, stochastic author whose output must be
quarantined until it has run (\fn{edit}, S12, confined by \law{D3}
and \law{D5}); a goal stated in language, versioned, and checked
by an executable predicate (\ty{State}, S15); requests that must
be recognized as repeats of earlier ones (\ty{EditCache}, S14);
and claims that must be kept distinct from results (\ty{Eval}
behind \law{D5}). Make has an author too, but a trusted human, and
that single assumption is why it needed none of this machinery.}
{Agent frameworks have
the author but let it execute and grade itself.}
{{\ifarxiv The full version quantifies\else Section~\ref{sec:prelim} quantifies\fi} what
breaks when either half is
used alone; the pairing, and the rules that make it safe, are the
contribution.}}

\ifarxiv
  \section{Status and Outlook}
\label{sec:outlook}

{%
This report is the first part of a longer effort. It fixed the
diagnosis (Section~\ref{sec:intro}), the eighteen requirements an
ADRS should guarantee and the database capabilities they mirror
(Table~\ref{tab:mapping}), and the two-stratum design with the five
rules from which the guarantees follow (Section~\ref{sec:model}).
The full version, in preparation, walks each requirement to its
guarantee on the running example: the reliable and non-wasteful
groups delivered by the executor's versioning, provenance,
incremental maintenance, and cost-based scheduling; transparency as
queries over the plan and the record; and collaboration, down to
peer review conducted as read-only queries over a submitted
instance. \revised{grammar: states -> state; visionary work -> vision}{We plan to also report early results from a prototype of the two strata and state the open problems the design exposes, from
semantic change impact to goal compilation. We welcome comments on this vision while the rest is under construction: the LLM should be the query compiler, never the executor.}}

\else
  \input{04_functionality}
  \input{05_performance}
  \input{06_transparency}
  \input{06_collaboration}
  \input{07_prelim}
  \input{08_agenda}
  \input{09_related}
  \input{10_conclusion}
\fi

\bibliographystyle{ACM-Reference-Format}
\bibliography{references}

\end{document}